\documentclass[onecolumn,showpacs,preprintnumbers,amsmath,amssymb]{revtex4}
\usepackage{graphicx}% Include figure files
\usepackage{dcolumn}% Align table columns on decimal point
\usepackage{bm}% bold math
\def\w6{{$\hat{W}_6$}}

\begin{document}

\title{Geometrical frustration in liquid Fe and Fe-based metallic glass}

\author{P. Ganesh}
\author{M. Widom}
\affiliation{Carnegie Mellon University\\
Department of Physics\\
Pittsburgh, PA  15213}

\date{\today}

\begin{abstract}
We investigate short range order in liquid and supercooled liquid Fe
and Fe-based metallic glass using {\em ab-initio} simulation methods.
We analyze the data to quantify the degree of local icosahedral and
polytetrahedral order and to understand the role of alloying in
controlling the degree of geometric frustration.  Comparing elemental
Fe to Cu~\cite{Cupaper} we find that the degree of icosahedral order
is greater in Fe than in Cu, possibly because icosahedral disclination
line defects are more easily incorporated into BCC environments than
FCC.  In Fe-based metallic glass-forming alloys (FeB and FeZrB) we
find that introducing small concentrations of small B atoms and large
Zr atoms controls the frustration of local icosahedral order.
\end{abstract}

\pacs{61.43.Dq,61.20.Ja,61.25.Mv}

\maketitle

\section{\label{sec:intro}Introduction}

As noted by Frank~\cite{Frank}, the local icosahedral clustering of 12
atoms about a sphere is energetically preferred because it is made up
entirely of four-atom tetrahedra, the densest-packed cluster possible.
However, local icosahedral order cannot be propagated throughout space
without introducing defects.  Frustration of packing icosahedra is
relieved in a curved space, where a perfect 12-coordinated icosahedral
packing exists~\cite{SM82,Nelson83,Sethna83}.  We adopt this structure
as an ideal reference against which actual configurations will be
compared.

Disclination line defects of type $\pm 72^{\circ}$ may be introduced
into this icosahedral crystal and thereby control the curvature.  In
order to ``flatten'' the structure and embed it in ordinary three
dimensional space an excess of $-72^{\circ}$ disclinations is needed,
and these cause increased coordination numbers of 14, 15 or 16.  Large
atoms, if present, would naturally assume high coordination numbers
and aid in the formation of a disclination line network. Similarly,
smaller atoms would naturally assume low coordination numbers of 8, 9
or 10, and have positive disclinations attached to them, increasing
the frustration. For a particular coordination number, it may be
possible to construct a cluster, known as a Kasper-Polyhedron (see
Section~\ref{sec:alloyW6}), made entirely of tetrahedrons.

Honeycutt and Andersen~\cite{Honeycutt} introduced a method to count
the number of tetrahedra surrounding an interatomic bond.  This number
is 5 for icosahedral order with no disclination, 6 for a $-72^{\circ}$
disclination and 4 for a $+72^{\circ}$ disclination.  Steinhardt,
Nelson and Ronchetti~\cite{Steinhardt} introduced the orientational
order parameter \w6 to demonstrate short range icosahedral order.  We
employ both methods to analyze icosahedral order in supercooled metals
and metal alloys, in addition to conventional radial distribution
functions, structure factors and Voronoi analysis.

Many simulations have been performed on pure elemental metals and
metal alloys using model
potentials,~\cite{Egami,Chen-Liu,Sadigh,Goddard}, but do not
necessarily produce reliable structures owing to their imperfect
description of interatomic interactions.  First principles ({\em
ab-initio}) calculations achieve the most realistic possible
structures, unhindered by the intrinsic inaccuracy of phenomenological
potentials and with the ability to accurately capture the chemical
natures of different elements and alloys.  The trade-off for increased
accuracy is a decrease in the system sizes one can study, so only
local order can be observed, not long range. Also runs are limited to
short time scales.  Early ab-initio studies on liquid
Copper~\cite{Vanderbilt,Hafner,Valladares} and Iron~\cite{Kresse} were
not analyzed from the perspective of icosahedral ordering.
Recent ab-initio studies on Ni and Zr~\cite{Jakse_Ni,Jakse_Zr} find
that the degree of icosahedral ordering increases with supercooling in
Ni, while in Zr BCC is more favored.  Studies on binary metal alloys
by Jakse {\em et al.}~\cite{jakse_alloy1,jakse_alloy2} and by Sheng
{\em et al}~\cite{MaNature} quantify local icosahedral order in the
alloys.  We previously~\cite{Cupaper} investigated icosahedral order
in liquid and supercooled Cu.

Elemental metals crystallize so easily that they can hardly be made
amorphous at any quench rate. Alloying can improve the ease of glass
formation. For some special alloys, a bulk amorphous state can be
reached by slow cooling. Pure elemental Fe is a poor glass former, but
Fe-based compounds like FeB and especially FeZrB, show improved glass
formability. We augment our molecular-dynamics simulation with another
algorithm called `Tempering' or `replica exchange method'
(REM)~\cite{REM,super_REM} for fast equilibration at low temperatures.

In comparison to liquid and supercooled liquid Copper~\cite{Cupaper}
which show only weak icosahedral order and very little temperature
variation, Fe showed a monotonic increase in icosahedral order, which
became very pronounced when supercooled.  Analysis of quenched Fe
revealed a natural way of introducing 5-fold coordinated bonds plus a
single -72$^{\circ}$ disclination line segment into an otherwise
perfect BCC environment, without disturbing the surrounding structure.
Addition of B to Fe decreased the icosahedral order, due to the
positive disclinations centered on the smaller B atom which increased
frustration. Further inclusion of larger Zr atoms to form FeZrB found
an enhanced icosahedral order compared to FeB.  This could possibly be
explained by formation of negative disclination line
defects~\cite{NelsonWidom} anchored on the larger Zr atoms, which
eases the frustration of icosahedral order on the Fe atoms.

At high temperatures all of our measured structural properties of
liquid Cu~\cite{Cupaper} and liquid Fe resembled each other, and also
resembled a maximally random jammed~\cite{MRJ} hard sphere
configuration.  This suggests that a nearly universal structure exists
for systems whose energetics are dominated by repulsive central
forces.

Section~\ref{sec:TMD} describes our combined method of monte-carlo and
first principles MD, that we refer to as ``Tempering MD'' and
discusses other simulation details. Section~\ref{sec:PureFe} presents
our results on pure Fe while (section~\ref{sec:alloy}) compares this
with FeB and FeZrB alloys.

\section{\label{sec:TMD}Tempering Molecular Dynamics (TMD) and other Simulation Details}

One reason alloys form glass more easily is that chemical identity
introduces a new configurational degree of freedom that evolves
slowly~\cite{Greer,Desre}. Unfortunately, this makes simulation more
difficult. It is especially difficult to equilibrate the system at
very low temperatures, because the probability to cross an energy
barrier drops, trapping it in particular configurations. For this
reason we use a Monte-Carlo method, known as tempering or replica
exchange~\cite{REM,super_REM} to augment our first-principles MD,
allowing us to sample the configurational space more efficiently than
conventional MD.

In the canonical ensemble, energy fluctuates at fixed temperature. A
given configuration $C$ with energy $E$ can occur at any temperature
$T$ with probability proportional to $e^{-\beta E}$,
($\beta=1/kT$). Now consider a pair of configurations, $C_{1}$ and
$C_{2}$ of energy $E_{1}$ and $E_{2}$ occurring in simulations at
temperatures $T_{1}$ and $T_{2}$.  We can take $C_{1}$ as a member of
the ensemble at $T_{2}$, and $C_{2}$ as a member of the ensemble at
$T_{1}$, with a probability
\begin{equation}
\label{eq:prob}
P=e^{-(\beta_2-\beta_1)(E_1-E_2)}  
\end{equation}
without disturbing the temperature-dependent probability distributions
of energy (or any other equilibrium property). Because each run
remains in equilibrium at all times even though its temperature
changes, we effectively simulate a vanishingly low quench rate.

In practice we perform several MD simulations at temperatures
separated by 100K. We use ultrasoft pseudopotentials~\cite{USPP} as
provided with VASP~\cite{USPP2} to perform the MD simulation. All
calculations are '$\Gamma$' point calculations (a single 'k' point).
All runs use an MD time step of 2 fs, and reach total simulated time
of order 1.5-1.8 ps (see Table~\ref{tab:alloys}) with a total of N=100
atoms. Every 10 MD steps we compare the energies of configurations at
adjacent temperatures and swap them with the above probability.
Eventually, configurations initially frozen at low temperature reach a
higher temperature. The simulations then can carry the structure over
energy barriers, after which the temperature can again drop.

\begin{table}
\caption{\label{tab:alloys}Details of tempering MD runs.}
\begin{tabular}{|r|r|r|r|}
\hline
Chemical Species & Temperatures(K) & density~(\AA$^{-3}$)  & time (ps)\\
\hline
Fe$_{100}$ & 800-1900 & 0.0756 & 1.5\\
\hline
Fe$_{80}$B$_{20}$ & 800-1500 & 0.0814 & 1.8 \\
\hline
Fe$_{70}$Zr$_{10}$B$_{20}$ & 800-1800 & 0.0787 & 1.8 \\
\hline
\end{tabular}
\end{table}

In an effort to explore the structures of compounds with differing
glass-forming ability we compare pure elemental Iron and two
Iron-based glass-forming alloys. Tempering MD requires that we perform
simulations at a constant density for all the temperatures, but we
have no rigorous means of predicting the density at high temperature.
For pure liquid Iron, the density is known
experimentally~\cite{Waseda}, and we use this value.  For FeB and
FeZrB, we took a high temperature liquid structure and quenched it,
relaxing positions and cell lattice parameters, to predict a low
temperature density.  We then decreased the density of the relaxed
structure by 6 percent to account for volume expansion, to arrive at
the densities used in our liquid simulations.

Because of the efficient sampling of our tempering MD method, the
structure of pure Fe partially crystallizes at low temperatures after
about 1 ps.  In the following discussion of our T=800K sample we will
refer to different structural features before and after
crystallization.  We also performed several long (2.0ps) conventional
first-principles MD at T=800K yielding results similar to the results
of tempering MD prior to crystallization.

For all runs we employed spin polarization, reasoning that local
magnetic moments exist even above the Curie point.  These local
moments have a significant influence on the short-range order because
ferromagnetic Iron prefers a longer bond length than paramagnetic
Iron~\cite{Kresse97}.  Of course, the ferromagnetic state of the
liquid implies improper long-range correlations.  Unfortunately, since
our forces are calculated for electronic ground states, we cannot
rigorously model the true paramagnetic state of liquid Iron and
Iron-based alloys with these methods.

\section{\label{sec:PureFe}Pure Fe}

\subsection{\label{sec:Fegr}Radial Distribution Function $g(r)$}

\begin{figure}
\includegraphics[width=3in,angle=-90]{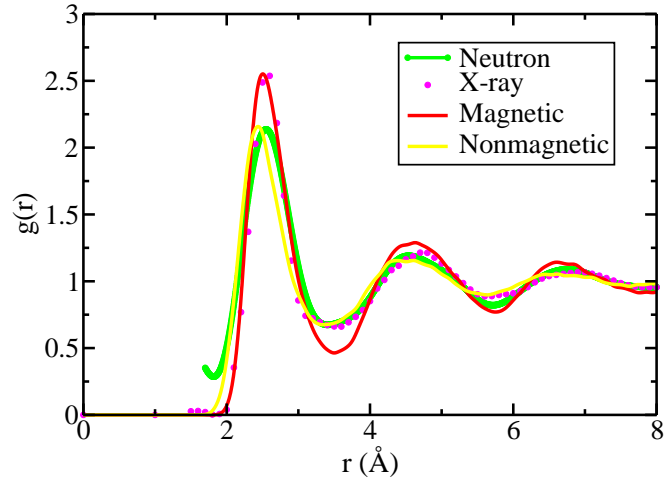}
\caption{\label{fig:Fegr}Radial distribution function of pure
elemental liquid Fe.  Simulations (magnetic and non-magnetic) are run
at T=1800K, compared with X-ray experiments at T=1833K~\cite{Waseda}
and Neutron scattering experiments at 1830K~\cite{Holland2}.}
\end{figure}

 The radial distribution function, $g(r)$, is proportional to the
 density of atoms at a distance $r$ from another atom and is
 calculated here by forming a histogram of bond lengths. We use the
 repeated image method to obtain the bond lengths greater than half
 the box size and anticipate $g(r)$ in this range may be influenced by
 finite size effects. Further, we smooth out the histogram with a
 gaussian of standard deviation 0.05\AA.

To evaluate the role of magnetism, Fig.~\ref{fig:Fegr} illustrates the
radial distribution functions for liquid Fe simulated at T=1800K, just
below melting (T$_m$=1833K).  Evidently, the simulation with magnetism
yields good agreement in the position and height of the first peak in
$g(r)$ with experimental X-ray $g(r)$~\cite{Waseda}, while neglect of
magnetic moments results in near neighbor bonds that are too short and
too weak.  However, magnetism overestimates the strength of long-range
correlations beyond the nearest-neighbor peak, while neglecting
magnetism yields reasonably accurate long-range $g(r)$. Nevertheless,
for the present study of local order, it is necessary to make spin
polarized calculations, to get the short range correlations and hence
the local order correct.  Strangely, a recent experimental neutron
$g(r)$~\cite{Holland2} has a shorter and broader first peak compared
to both our magnetic $g(r)$ and to the $g(r)$ from the prior X-ray
diffraction experiment~\cite{Waseda}. The positions of the different
maxima and minima in our simulated magnetic $g(r)$ compare well
with both the experiments. The position of the first peak in our
magnetic $g(r)$ is shifted by 0.05\AA~to the left of the neutron first
peak. The X-ray experiment doesn't have enough data points around the
first maximum to determine the peak position accurately.

%Experimental $g(r)$'s are obtained by fourier transform of the
%experimentally measured liquid structure factor $S(q)$
%(Section~\ref{sec:sq}). In the following subsection we show that the
%recently obtained neutron $S(q)$ does not obey a sum
%rule~\cite{Norman,Krogh-Moe} relating to vanishing $g(r)$ at short
%distances, indicating a possible difficulty in normalizing or
%background subtraction. We also show that our $S(q)$, obtained from
%the simulated $g(r)$ obeys the sum rule.

We calculate the coordination number by counting the number of atoms
within a cutoff distance from a central atom. We choose the cutoff
distance ($R_{cut}$) at the first minimum of $g(r)$. For pure Fe the
minimum is at $R_{cut}$=3.5~\AA. The precise location of the minimum
is difficult to locate, and its variation with temperature is smaller
than the error in locating its position, so that we don't change the
value of $R_{cut}$ with temperature. With this value of $R_{cut}$ we
find an average coordination number ($N_c$) of 13.2 which is nearly
independent of temperature ($N_c$ changes from 13.1 at high
temperature to 13.3 with supercooling).

\subsection{\label{sec:sq} Liquid Structure Factor S(q)}

The liquid structure factor $S(q)$ is related to the radial
distribution function $g(r)$ of a liquid with density $\rho$ by,
\begin{equation}
\label{eq:Sqeq}
S(q)=1+4 \pi \rho \int\limits_{0}^{\infty}[g(r)-1] {\sin(qr)\over qr}r^2dr
\end{equation}
One needs the radial distribution function up to large values of $r$
to get a good $S(q)$. In our first principles simulation, we are
restricted to small values of $r$, due to our small system sizes, so
we need a method to get $S(q)$ from our limited $g(r)$ function.
Baxter developed a method~\cite{Baxter,Jolly} to extend $g(r)$ beyond
the size of the simulation cell. The method exploits the short range
nature of the direct correlation function $c(r)$, which has a range
similar to the interatomic interactions~\cite{Ashcroft}, as opposed to
$g(r)$ which is long ranged.
%The exact relation that connects
%these two functions is the Ornstein-Zernike relation,
%\begin{equation}
%\label{eq:O-Z}
%h(r)=c(r)+\rho \int h(|{\bf r-r'}|)c(|{\bf r'}|)d{\bf r'}
%\end{equation}  
%where $h(r)=g(r)-1$.

Assuming that $c(r)$ vanishes beyond a certain cutoff distance $r_c$,
%\begin{eqnarray}
%rc(r)=-Q'(r)+2\pi \rho{\int \limits_{r}^{r_c} Q'(r')Q(r'-r)dr'}
%\label{cr}
%\end{eqnarray}
%and
%\begin{eqnarray}
%rh(r)=-Q'(r)+2\pi \rho {\int\limits_{0}^{r_c}(r-r')h(|r-r'|)Q(r')dr'}.
%\label{hr}
%\end{eqnarray}
%Here $Q(r)$ is zero for $r> r_c$, and continuous at $r_c$, and
%$Q'(r)=dQ(r)/dr$. The continuity
%of $Q(r)$ means that,
%\begin{equation}
%Q(r)=-{\int \limits_{r}^{r_c}dr' Q'(r')}.
%\label{qr}
%\end{equation} 
%The remarkable property of this method is that if we know $h(r)$ over
%a range $0\le r \le r_c$, then we can obtain $c(r)$ over its entire
%range (from 0 to $r_c$), which implicitly determines $h(r)$ over its
%entire range (from 0 to $\infty$) through Eq.~(\ref{eq:O-Z}).
we solve the Baxter's equations iteratively to obtain the full direct
correlation function for $0<r<r_c$. From $c(r)$ we calculate the
structure factor $S(q)$ by a standard Fourier Transform.
%\begin{equation}
%S(q)={1\over 1-\rho \hat c(q)}
%\label{OZink}
%\end{equation}
%where,
%\begin{equation}
%\label{eq:cq}
%\hat c(q)=4 \pi \int \limits_{0}^{\infty}r^2 c(r) {\sin(qr)\over qr}dr. 
%\end{equation}
The $S(q)$ showed good convergence with different choices of $r_c$.  A
choice of $r_c$=5\AA~ seemed appropriate because it was one half of
our smallest simulation cell edge length. Even though in metals there
are long range oscillatory Friedel oscillations, our ability to
truncate $c(r)$ at $r_c$=5\AA~ shows that these are weak compared with
short range interactions. An application of this method to obtain
$S(q)$ of Cu~\cite{Cupaper} showed excellent agreement with the
experimental $S(q)$.

\begin{figure}
\includegraphics[width=3in,angle=-90]{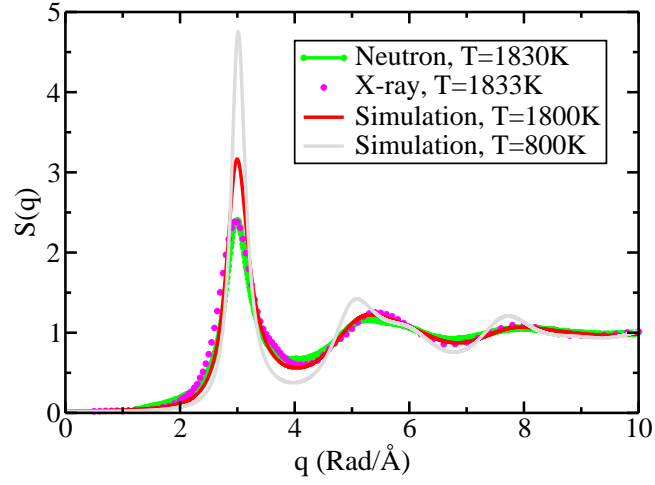}
\caption{\label{fig:Fesq} Comparison of simulated and experimental S(q) near the melting temperature of Fe.} 
\end{figure}

The simulated $S(q)$ for pure Fe at T=1800K (see Fig.~\ref{fig:Fesq})
is compared to recent neutron scattering experiments at T=1830K. Even
though the positions of the different peaks compare very well, there
is serious discrepancy in their heights. Especially, the height of the
first peak of our simulated $S(q)$ is higher than that of the
experiments.  This discrepancy is expected because we include
magnetism, which gives accurate short-range correlations while
overestimating the long-range ones (see Fig.~\ref{fig:Fegr}).  But the
cause of the discrepency is not entirely clear since a comparison of
the simulated structure factor of Ni~\cite{Jakse_Ni} (done without
including magnetism) with neutron scattering
experiments~\cite{Holland2} shows similar discrepancies between their
$S(q)$'s.

A sum rule can be obtained for $S(q)$~\cite{Norman,Krogh-Moe}. By
inverting the fourier transform of Eq.~\ref{eq:Sqeq} and then taking
the $r \rightarrow 0$ limit, one gets
\begin{equation}
\label{SumRule}
I(Q) \equiv \int \limits_{0}^{Q} q^2[S(q)-1]dq\rightarrow-2\pi^2\rho
\end{equation}
in the limit $Q\rightarrow \infty$.  Further, the integral is supposed
to oscillate with $Q$ about the limiting value as $Q \rightarrow
\infty$. Using our $S(q)$ we observed that the integral is consistent
with the sum rule and oscillates nicely about the limiting value for
$Q \ge 3 $\AA$^{-1}$, while using the $S(q)$ from the neutron
scattering experiments~\cite{Holland2}, we observe a positive drift in
the mean value about which the integral oscillates. Such a drift could
indicate the presence of spurious background corrections.  The $S(q)$
from the X-ray experiment~\cite{Waseda} seems to be in good agreement
with the ideal sum rule.

As we lower the temperature, the peak heights in $S(q)$ grow,
indicating an increase in short range order with supercooling. We also
observe a slight shoulder in the second peak of the $S(q)$
(Fig.~\ref{fig:Fesq}), which grows with supercooling.  The split
second peak positions are in the ratios of 20:12 and 24:12 with
respect to the first peak positions, just what one would ideally
observe if there was icosahedral order~\cite{NelsonWidom,Sachdev}.

\subsection{\label{sec:alloyW6} Bond Orientation Order Parameter \w6}

%\begin{table}
%\caption{\label{tab:W6} $\hat{W}_6$ values for a few clusters.  The
%  Z8-Z16 series are Kasper-Polyhedrons including the perfect
%  icosahedron (Z12).}
%\begin{tabular}{|c||c|c|c||c|c|c|c|c|c|c|}
%\hline

%Cluster & HCP & FCC & BCC & Z8 &Z9&Z10&Z12&Z14&Z15&Z16\\
%\hline
%No. of atoms & 12 & 12 & 14 & 8 &9 &10 &12&14&15&16\\
%\hline 
%Voronoi type &
%(0,12,0)&(0,12,0)&(0,6,0,8)&(0,4,4)&(0,3,6)&(0,2,8)&(0,0,12)&(0,0,12,2)&(0,0,12,3)&(0,0,12,4)\\
%\hline
%$\hat{W}_6$ &  -0.012 & -0.013 & +0.013& +0.010 &-0.038 &-0.093&-0.169&-0.093&-0.037&+0.013\\
%\hline
%\end{tabular}
%\end{table}

\begin{table}
\caption{\label{tab:W6} $\hat{W}_6$ values for a few clusters.  The
  Z8-Z16 series are Kasper-Polyhedrons including the perfect
  icosahedron (Z12).}
\begin{tabular}{|c||c|c|c|c|}
\hline
Cluster &No. of atoms &Voronoi type&$\hat{W}_6$\\
\hline
HCP&12&(0,12,0)&-0.012\\
FCC&12&(0,12,0)&-0.013\\
BCC&14&(0,6,0,8)&+0.013\\
Z8&8&(0,4,4)&+0.010\\
Z9&9&(0,3,6)&-0.038\\
Z10&10&(0,2,8)&-0.093\\
Z12&12&(0,0,12)&-0.169\\
Z14&14&(0,0,12,2)&-0.093\\
Z15&15&(0,0,12,3)&-0.037\\
Z16&16&(0,0,12,4)&+0.013\\
\hline
\end{tabular}
\end{table}

\begin{figure}
\includegraphics[width=3in,angle=-90]{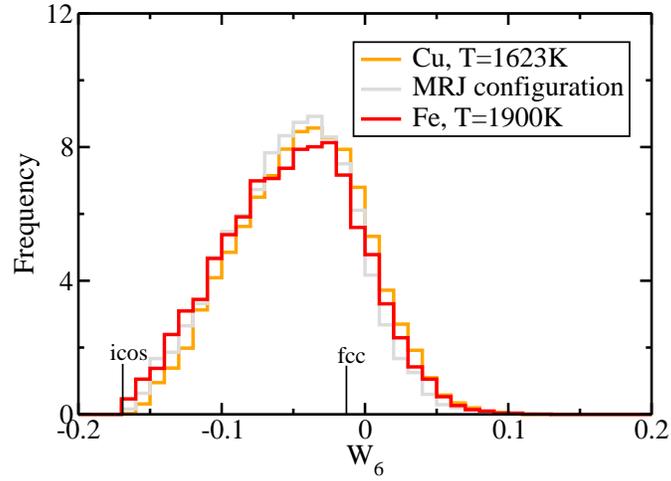}
\caption{\label{fig:w6_Fe_Cu}Distribution of \w6 in liquid and supercooled liquid Fe and Cu. Fe shows more pronounced icosahedral order than Cu with supercooling}
\end{figure}

\begin{figure}
\includegraphics[width=3in,angle=-90]{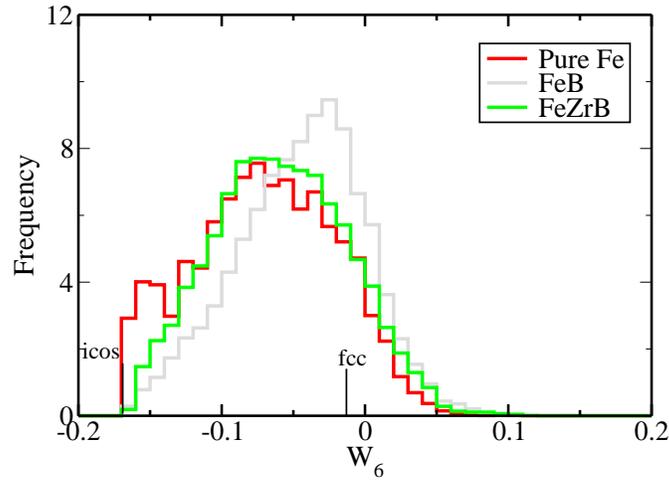}
\caption{\label{fig:Fe_w6lowT}Distribution of \w6 around Fe atoms, in
supercooled Fe, FeB and FeZrB at T=800K}
\end{figure}

Steinhardt, {\em et al.}~\cite{Steinhardt} introduced the $\hat{W}_l$
parameters as a measure of the local orientational order in liquids
and undercooled liquids. To calculate $\hat{W}_l$, the orientations of
bonds from an atom to its neighboring atoms are projected onto a basis
of spherical harmonics.  Rotationally invariant combinations of
coefficients are then averaged over many atoms in an ensemble of
configurations. The resulting measures of local orientational order
can be used as order parameters to characterize the liquid
structures. For an ideal icosahedral cluster, $l=6$ is the minimum
value of $l$ for which \w6 $\ne 0$. Table~\ref{tab:W6} enumerates \w6
values for different ideal clusters. The ideal icosahedral value of
\w6 is far from other clusters, making it a good icosahedral order
indicator.

Kasper-Polyhedrons~\cite{Kasper_Doye,Nelson83,Frank-Kasper} are
polyhedrons which minimize the number of disclinations for a
particular coordination number.  The series Z8-Z16 in
Table.~\ref{tab:W6} are such Kasper-Polyhedrons but with the added
constraint that the surface atoms be triangulated with equilateral
triangles.  The icosahedron with a coordination of 12 is one such
Kasper-Polyhedron with no disclinations.  Adding disclinations to the
icosahedron, one finds that each disclination increases the \w6 value
by the same amount irrespective of its sign.

As before, we choose the cutoff distance to specify near neighbors as
$R_{cut} =3.5$\AA.  For pure Fe at high temperatures, \w6 resembles
that of high temperature liquid Cu~\cite{Cupaper}, which in turn
resembles the maximally random jammed configuration of hard
spheres~\cite{MRJ}(Fig.~\ref{fig:w6_Fe_Cu}). On this basis we suggest
that the MRJ configuration represents an idealized structure that is
universal for strongly repulsive interactions. All pure metallic
systems might approach this ideal structure at sufficiently high
temperature.

However, as temperature drops the \w6 distribution shifts strongly to
the left, with a pronounced increase in \w6 $\leq$ -0.1
(Fig.~\ref{fig:Fe_w6lowT}). This indicates a rather high concentration
of nearly icosahedral clusters in supercooled liquid Fe.  Even at very
low percentages of supercooling of Cu ($\sim$ 3\%~\cite{Cupaper}) and
Fe ($\sim$ 2\%, not shown), Fe shows a clear enhancement in the
negative \w6 distribution as compared to Cu.

We checked the \w6 distribution of a non-magnetic simulation to see if
it is strongly influenced by magnetism, and found a nearly identical
result.  In particular, we still found a strong enhancement of the
nearly icosahedral clusters relative to liquid Cu or pure Fe at high
temperature.

\subsection{\label{sec:Vor}Voronoi Analysis}

\begin{table}
\caption{\label{tab:Voronoi} Voronoi statistics (\% of Fe atoms) for typical Fe-centered clusters in Fe, FeB and FeZrB. Edges smaller than 0.33\AA~ and faces smaller than 0.3~\AA$^2$ have been treated as a single vertex.}
\begin{tabular}{|c|c|c|c|c|c|c|}
\hline
Voronoi&\multicolumn{4}{c}{Fe}
&\multicolumn{1}{|c}{FeB}
&\multicolumn{1}{|c|}{FeZrB}\\
\cline{2-7}
type&\multicolumn{2}{c|}{supercooled} 
%&\multicolumn{1}{c|}{}
&\multicolumn{2}{c|}{defective} 
&\multicolumn{1}{c|}{supercooled}
&\multicolumn{1}{c|}{supercooled}\\
%&\multicolumn{1}{c|}{} && \\
&\multicolumn{2}{c|}{liquid} 
%&\multicolumn{1}{c|}{}
&\multicolumn{2}{c|}{crystal} 
&\multicolumn{1}{c|}{liquid}
&\multicolumn{1}{c|}{liquid}\\
%&\multicolumn{1}{c|}{} && \\
\cline{2-7}
&\multicolumn{1}{c|}{800K}
&\multicolumn{1}{c|}{relaxed}
&\multicolumn{1}{c|}{800K}
&\multicolumn{1}{c|}{relaxed}
&\multicolumn{1}{c|}{800K} 
&\multicolumn{1}{c|}{800K}\\
\hline
(0,0,12) &8.3 &11.0 &5.0 &0.0 &0.6 &1.4\\
(0,0,12,2) &2.1 &3.6 &2.0 &2.0 &0.5 &0.3\\
(0,2,10,\{0,1\}) &6.2 &6.8 &2.0 &0.0 &4.4 &7.1\\
(0,1,10,2) &3.6 &5.4 &3.0 &0.0 &2.1 &3.1\\
(0,4,8,\{0,1,2,3\}) &7.9 &4.8 &10.0 &6.0 &8.6 &9.4\\
(0,3,8,\{0,1,2,3\}) &6.8 &6.0 &12.0 &0.0 &4.3 &8.3\\
(0,2,8,\{1,2,3,4\}) &3.0 &3.0 &3.0 &0.0 &0.0 &0.0\\
(0,3,6,4) &2.1 &2.2 &2.0 &6.0 &0.0 &0.0\\
(0,5,4,4) &1.5 &1.2 &3.0 &18.0 &0.0 &0.0\\
(0,6,0,8) &0.0 &0.4 &0.0 &56.0 &0.0 &0.0\\
\hline
\end{tabular}
\end{table}

To explain the origin of this low \w6 peak in pure Fe, we performed a
Voronoi analysis~\cite{Finney} of the liquid before and during
crystallization.  A Voronoi polyhedron is described by indices
$(F_{3},F_{4},F_{5},...)$ where $F_{i}$ denotes the number of faces
with $i$ edges.  For example (0,0,12) denotes an icosahedron, while
(0,0,12,2) denotes a 14-coordinated (Z14) atom, with 12 5-fold bonds
and 2 6-fold bonds.  The (0,0,12,2) is a characteristic TCP
(tetragonal close-packed) structure of the Frank-Kasper type, with a
$-72^{\circ}$ disclination line running through an otherwise perfect
icosahedron.  In a body-centered cubic crystal all atoms are of
Voronoi type (0,6,0,8), which is an alternate 14-coordinated
structure.

Supercooled liquid Fe at T=800K (Table~\ref{tab:Voronoi}) contains a
high fraction of icosahedral atoms of type (0,0,12) and (0,0,12,2)
. Those with Voronoi type (0,2,10,\{0,1\}) and (0,1,10,2) also have
very negative \w6, so that they can be thought of as related to the
icosahedron.  Together, they explain the enhanced negative \w6
distribution. Supercooled Cu (which shows weak icosahedral order) and
high temperature Fe (T=1900K), in contrast contain about 1.2\% of
(0,0,12) and no (0,0,12,2). The MRJ configuration contains 1.2\%
(0,0,12) but no (0,0,12,2).

Strikingly, the icosahedral clusters tend to join in pairs and strings
of 3 atoms in length.  An instantaneous quench of a particular liquid
structure at 800K which had a high fraction of icosahedral units,
using conjugate gradient relaxation of the atomic coordinates and
lattice parameters, shows a clear enhancement of the icosahedral and
other closely related units. The strings of icosahedral units found in
the supercooled liquid connected to form networks.  A quench starting
from a different instantaneous supercooled liquid structure at 800K
containing fewer icosahedral structures resulted in rapid
crystallization. This indicates that the presence of icosahedrons may
inhibit crystallization.  Similar quenches, starting from
instantaneous liquid structures at higher temperatures, also resulted
in relaxed structures with some of them showing a high negative \w6
distribution comparable to the distribution at 800K.  The quenches
that partially crystallized showed a high fraction of BCC (0,6,0,8)'s
($\sim$ 40\%) and was always accompanied by a high fraction of
(0,5,4,4)'s ($\sim$ 25\%), which are otherwise absent or very low in
the liquid.

Under TMD the supercooled liquid at T=800K eventually
crystallizes, with the \w6 distribution peaked strongly around zero
consistent with the value for ordinary crystalline clusters, but
retaining a subset of atoms with nearly icosahedral \w6$<$-0.14.  Not
surprisingly, these were precisely the atoms that had Voronoi type
(0,0,12) prior to crystallization. The Z14 (0,0,12,2) atoms (with
\w6 $\sim$ -0.093) were mutual near-neighbors, linked along their
6-fold ($-72^{\circ}$) bonds, and also were neighbors of the nearly
icosahedral atoms.

\begin{figure}
\includegraphics[width=3in]{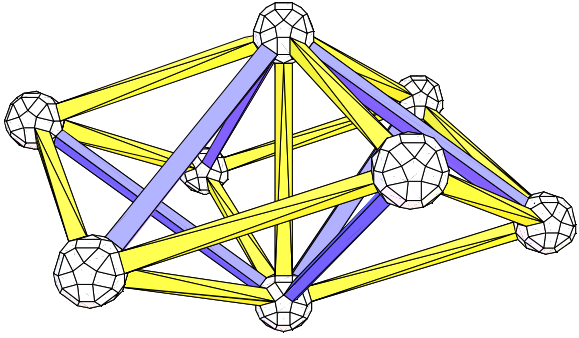}
\caption{\label{fig:defect}Fragment of BCC crystal illustrating
a central bond in the [1,1,1] (vertical) direction, surrounded by two
equilateral triangles rotated 60$^{\circ}$ from each other and
displaced on either side of the midplane.  Balls and sticks
follow ZomeTool convention: Balls are icosahedral, yellow sticks are
3-fold and blue sticks are 2-fold bonds.}
\end{figure}

We quenched this sample by conjugate gradient relaxation of atomic
coordinates and lattice parameters.  After relaxation we found 56
atoms had BCC Voronoi type (0,6,0,8).  Six icosahedral atoms became
12-coordinated (0,4,8) structures surrounding the bond connecting the
two (0,0,12,2) disclinated icosahedral atoms, which retained their type.
The remaining atoms served to link the cluster of icosahedron-related
atoms to the surrounding defect-free BCC crystal.

The icosahedron-related structure thus forms a point defect in an
otherwise perfect BCC crystal.  A simple way to create this defect is
to take two consecutive triangles surrounding a near-neighbor bond
along the BCC [1,1,1] direction (see Fig.~\ref{fig:defect}), displace
them into the perpendicular bisecting plane and rotate by
30$^{\circ}$.  The atoms along the [1,1,1] bond are now connected to
each other by a six-fold bond ($-72^\circ$ disclination) and are
connected to the six displaced atoms by five-fold bonds. This scheme
to transform a BCC rhombic dodecahedron into a Frank-Kasper Z14
polyhedron was described in Ref~\cite{Dubois,Khmelevskaya} in an
attempt to explain certain diffraction anamolies in iron and
vanadium-based alloys under ion irradiation.

Manually removing the defect, by reversing the above procedure then
relaxing, yields a perfect BCC crystal with all atoms in a (0,6,0,8)
Voronoi environment.  We also embedded the point defect in an
otherwise perfect BCC crystal with 128 atoms in a cubic box.
Relaxation showed that the defect was stable. The energy of the defect
was 6.18eV and the fractional volume increase was 0.013. 

Since tungsten, like Iron, crystallizes in BCC, we performed a
separate first-principles simulation of liquid Tungsten.  After
supercooling by 13\% to T=3200K, we found the \w6 histogram resembled
that of Fe at 1600K (which is 15\% supercooled).  A Voronoi analysis
revealed a similar percentage of (0,0,12)'s and (0,0,12,2)'s in both
these elemental BCC-forming metals. Hence we believe the relationship
between BCC and icosahedral structures may be linked to our observed
high fraction of icosahedra in liquid Fe as compared to Cu.  This may
also explain the reason why we are able to supercool BCC Fe more
deeply than FCC Cu in our simulations.

We also included an icosahedron point defect inside an otherwise
perfect FCC crystal of Cu, with 256 atoms in a cubic box. Relaxation
showed that the defect was stable. The energy cost of the defect was
4.69eV and the fractional volume increase was 0.012. Even though
this FCC defect cost less energy than the point defect in BCC, and
also needs little rearrangement of atoms as reflected by the slightly
lower fractional density change, we do not see a significant
icosahedral order in liquid Cu, when compared to liquid Fe. This could
be because the number of icosahedrons that a single Z14 disclination
can stabilize (up to six) is greater than one. So even an equal number
of the two different defects in BCC and FCC would result in more
icosahedral order in BCC than in FCC.

\subsection{\label{sec:Fe_HA}Honeycutt and Andersen analysis} 

\begin{figure}
\includegraphics[width=3in,angle=-90]{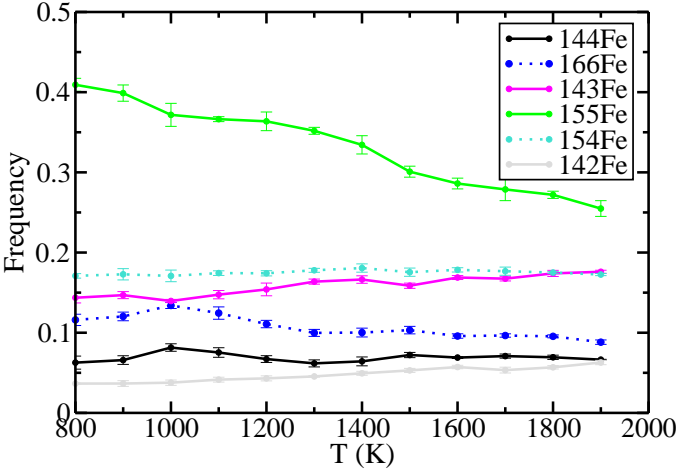}
\caption{\label{fig:HA_Fe}Honeycutt-Andersen analysis for pure Fe
shows a clear increase in five-fold bonds with supercooling.}
\end{figure}

Honeycutt and Andersen~\cite{Honeycutt} introduced a useful assessment
of local structure surrounding interatomic bonds.  We employ a
simplified form of their analysis, counting the number of common
neighbors shared by a pair of near-neighbor atoms.  This identifies
the number of atoms surrounding the near-neighbor bond and usually
equals the number of edge-sharing tetrahedra whose common edge is the
near-neighbor bond. We assign a set of three indices to each bond. The
first index is 1 if the root pair is bonded (separation less than or
equal to $R_{cut}$). The second index is the number of near-neighbor
atoms common to the root pair, and the third index gives the number
of near-neighbor bonds between these common neighbors. We take the
same value of $R_{cut}$=3.5\AA~ as mentioned before. Note that the Honeycutt
and Andersen fractions depend sensitively on $R_{cut}$, making precise
quantitative comparisons with other prior studies difficult.

In general, 142's are characteristic of close packed structures (FCC
and HCP) and 143's are characteristic of distorted
icosahedra~\cite{Ni-Ag}.  They can also be considered as
$+72^{\circ}$ disclinations~\cite{SM82,Nelson83,Sethna83}.  Likewise,
15's are characteristic of icosahedra, with 155's characterizing
perfect icosahedra while 154's and 143's characterize distorted
ones. 16's indicate $-72^{\circ}$ disclinations.  166's and 144's are
also characteristic of BCC.

 A Honeycutt and Andersen analysis for pure Fe, with an
$R_{cut}$=3.5\AA~ (Fig.~\ref{fig:HA_Fe}), showed that with
supercooling the fraction of 15 bonds rises from 0.46 at T=1900K to
0.59 at T=800K in the liquid before crystallization. The fraction of
155's (characteristic of perfect icosahedra) was always larger than
154's (characteristic of distorted icosahedra), and seemed to be
steeply increasing with supercooling as opposed to 154's, which were
relatively flat. The fraction of 14 bonds drop from 0.32 to 0.30, with
the icosahedral 143's being always higher than the cubic 142's. The
144's, which are characteristic of BCC remain nearly flat, even though
the 166's show a slight increase.  The ease of embedding a Z14
disclination in BCC Fe (see section~\ref{sec:Vor}) might explain the
slight increase in the 166's.

\section{\label{sec:alloy} Fe-B and Fe-Zr-B}
\subsection{\label{sec:Alloygr}Radial Distribution Function $g(r)$}

\begin{figure}

\includegraphics[width=3in,angle=-90]{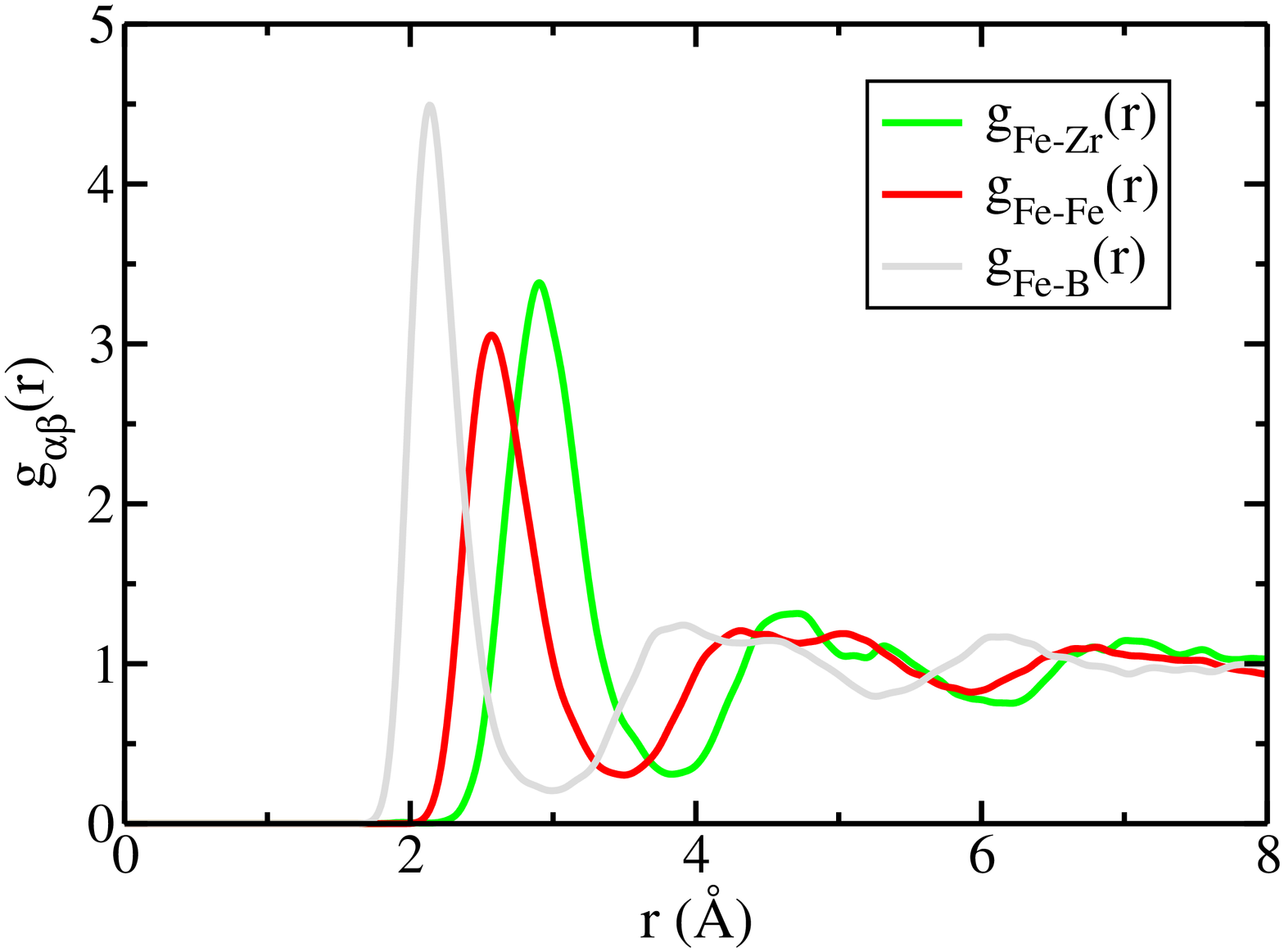}
\includegraphics[width=3in,angle=-90]{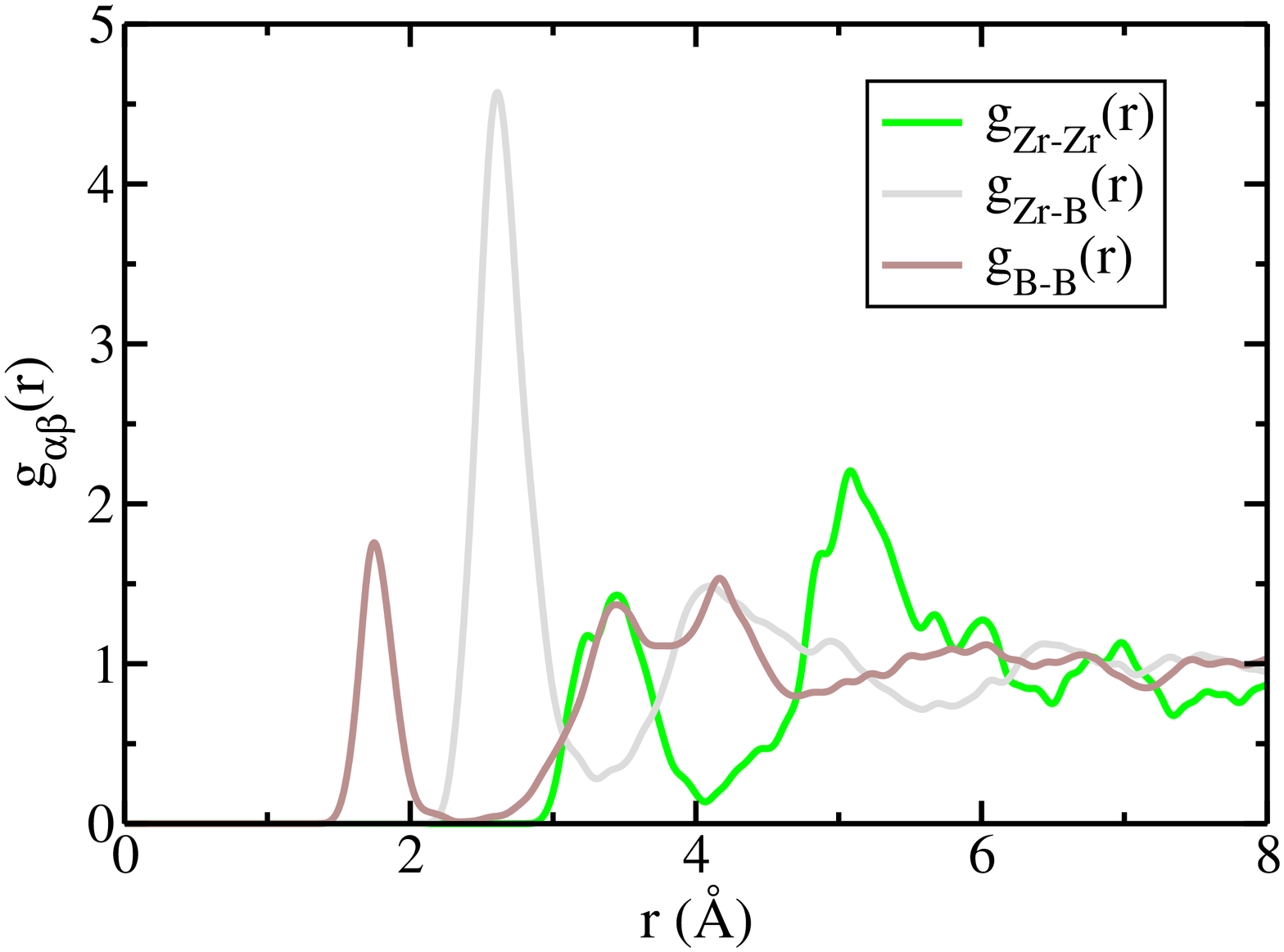}
\caption{\label{fig:Alloy_gr}Partial radial distribution functions in FeZrB at T=800K.}
\end{figure}

Fig.~\ref{fig:Alloy_gr} shows the pair correlation functions of
supercooled FeZrB at T=800K. From the heights of the first peaks, we
see that the strongest bonds form between the metalloid (B) and the
metal (Fe or Zr). The relative bond lengths reveal, as expected, that
B behaves as a small atom, Fe is medium sized and Zr is
large. Comparing simulations of FeZrB to pure Fe, we find that
$g_{FeFe}$ is reduced by alloying with B or ZrB, because Fe prefers to
associate with B or Zr rather than with Fe.

\begin{table}
\caption{\label{tab:NcFeB}Average coordination number in FeB.}
\begin{tabular}{|r|r|r|r|r|}
\hline
&FeFe & FeB & BFe & BB\\
\hline
$R_{cut}$&3.4&3.0&3.0&2.3\\
\hline
$N_{\alpha\beta}$&12.0 & 2.2 & 8.9 & 0.4 \\
\hline
\end{tabular}
\end{table}

\begin{table}
\caption{\label{tab:NcFeZrB}Average coordination number in FeZrB.}
\begin{tabular}{|r|r|r|r|r|r|r|r|r|r|}
\hline
&FeFe& FeZr & FeB & ZrFe & ZrZr &
ZrB & BFe & BZr & BB\\
\hline
$R_{cut}$&3.4&3.8&3.0&3.8&4.0&3.3&3.0&3.3&2.4\\
\hline
$N_{\alpha\beta}$&9.5 & 2.0 & 2.0 & 14.0 & 0.9 & 2.8 & 6.8 & 1.4 & 0.4 \\
\hline
\end{tabular}
\end{table}

To calculate the coordination number, define $N_{\alpha \beta}$ as the
average number of atoms of type $\beta$ around an atom of type
$\alpha$, We set $R_{cut}$ at the first minima of the partial radial
distribution functions (Fig.~\ref{fig:Alloy_gr}).  We list the partial
coordination numbers of FeB in Table.~\ref{tab:NcFeB} and FeZrB in
Table.~\ref{tab:NcFeZrB} (averaged over all the temperatures since the
temperature dependence is very weak and non-monotonic). The average
value of $N_{FeFe}$ decreases with alloying, due to decrease in the
concentration of Fe and the preference to bind with B and Zr. Zr,
being a large atom, has a larger coordination number.  Also note that
B and Zr favor each other more than themselves. We do find some B-B
pairs in the liquid state. 
\subsection{\label{sec:Alloysq} Liquid Structure Factor S(q)}

\begin{figure}
\includegraphics[width=3in,angle=-90]{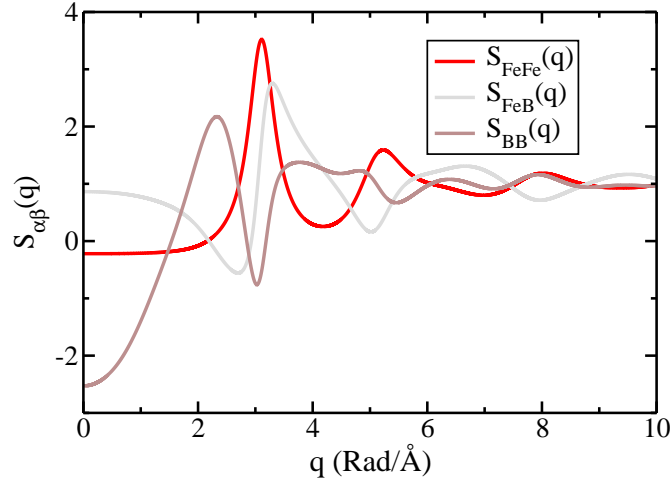}
\caption{\label{fig:Sq_FeB}Simulated partial structure factors of
 FeB at T=800K}
\end{figure}

Fig.~\ref{fig:Sq_FeB} shows the Faber-Ziman~\cite{Fischer} partial
structure factors of FeB at the lowest simulated temperature of
T=800K, defined as,
\begin{equation}
%S(q)=\sum_{\alpha\beta} x_\alpha x_\beta S_{\alpha\beta}(q)
S_{\alpha\beta}(q)=1+4\pi\rho\int\limits_{0}^\infty[g_{\alpha\beta}(r)-1] {\sin(qr)\over{qr}}dr
\end{equation}
The positions of the first and second peaks in partial $S_{FeFe}(q)$
is in very good agreement with the experimental results for amorphous
FeB~\cite{Lamparter}. Also, at the position of the splitting of the
second peak in $S_{FeFe}(q)$, as observed in the experiments, we
observe a slight shoulder.  Similarly, the positions of the different
peaks in $S_{FeB}(q)$ and $S_{BB}(q)$ are in agreement with the
experiments.

For FeB, the $q \rightarrow 0$ limit of $S_{FeFe}(q)$ is comparable to
the experimental value. The $q \rightarrow 0$ limit in the other two
partials differ from the experiment, $S_{BB}(q)$ more seriously than
$S_{FeB}(q)$. We think that this discrepancy in the long wavelength
regime is probably due to the very low density of B in our system
leading to poor statistics. Nevertheless, the excellent agreement in
the positions of the different peaks in the partial structure factors
shows that we have reasonably good representative structures of FeB at
T=800K.  Partial structure factors or partial pair distribution
functions are not available experimentally to compare with FeZrB
simulations.

\subsection{\label{sec:AlloyW6} Bond Orientation Order Parameter \w6}

\begin{figure}
\includegraphics[width=3in,angle=-90]{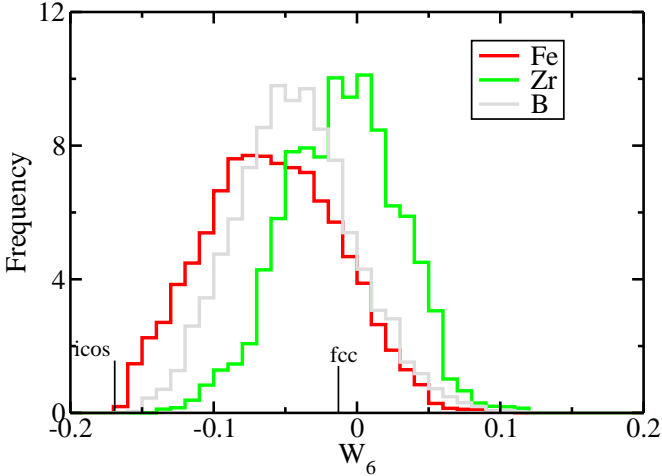}
\caption{\label{fig:otherw6}Distribution of \w6 for different chemical
species in supercooled FeZrB at T=800K}
\end{figure}

To define the \w6 distribution in an alloy, we concentrate on central
atoms of some particular species (e.g. Fe) but consider the
neighboring atoms of all species.  We chose the near-neighbor cutoff
distances as before.  Compare the Fe-based \w6 distributions at their
supercooled temperatures in Fig.~\ref{fig:Fe_w6lowT}.
% The FeB distribution is similar to high temperature \w6 distribution
%of Fe, but the FeZrB curve has significantly shifted to the left,
%extending more towards the ideal icosahedral value of \w6.  
The origin of high negative \w6 values for pure Fe was previously
explained in section~\ref{sec:Vor}.  Replacing a few medium sized Fe
atoms with smaller B atoms causes negative disclination lines to
concentrate on Fe, leading to a drop in the ideal icosahedral
clustering on these Fe atoms and strongly reducing the extreme
negative values of \w6 in FeB.  On the contrary, inclusion of Zr in
FeZrB causes negative disclinations to attach to Zr, easing
frustration, leading to more Fe centered clusters with icosahedral
ordering and increasing the negative region of \w6, as compared to
FeB. Inclusion of the big Zr atom enhances icosahedral order on the Fe
atoms.
 
Fig.~\ref{fig:otherw6} shows the \w6 distributions for FeZrB with
centers at Fe, Zr and B. The histogram with the center at Zr is almost
symmetric about the value of zero.  This suggests that the local
environment about Zr atoms is nearly spherical, as is expected given
its large size.  The B centered \w6 histogram is also asymmetric
towards negative \w6 values due to Kasper-Polyhedrons and slightly
distorted versions of them (see Table.~\ref{tab:W6} and
Section~\ref{sec:alloyW6}).

\subsection{\label{sec:AlloyVor}Voronoi Analysis}

%\begin{table}
%\caption{\label{tab:Voronoi_alloy}Voronoi analysis for supercooled FeB
%and FeZrB at T=800K}
%\begin{tabular}{|c||c|c|}
%\hline
%Fe centred Voronoi& percentages in FeB &percentages in FeZrB \\
%\hline
%(0,0,12)&0.5 &1 \\     
%(0,0,12,2)& 0.4&0.2\\   
%(0,2,10,\{0,1,2\})& 3.5&5\\ 
%(0,1,10,2)& 1.7 &2.2\\         
%(0,4,8,\{0,1,2,3,4\})& 6.9&6.6\\ 
%(0,3,8,\{1,2,3,4\})& 4.4&5.8\\
%\hline
%\hline
%B centred Voronoi & percentages in FeB&percentages in FeZrB\\
%\hline
%(0,3,6)&3.2&3.6\\
%(0,3,6,1)&1.4&0.4\\
%(0,5,4)&4.0&4.8\\
%(0,4,4)&0.6&1.2\\
%(0,2,8)&2.0&1.4\\
%\hline
%\end{tabular}
%\end{table}

A Voronoi analysis was performed for FeB in the supercooled liquid at
T=800K (Table~\ref{tab:Voronoi}). The Fe environments were mostly
(0,4,8,x) or (0,3,8,x) types, where x=\{0,1,2,3,4\}, with the higher
coordination polyhedron being more favored.  This was followed by the
very negative \w6 valued (0,0,12), (0,2,10,x)'s and (0,1,10,2)'s
Voronoi types occurring at lower frequency than in pure Fe.  Boron
mainly had environments of type (0,3,6)'s ($\sim$ 15\% of B atoms)
(Kasper polyhedron for Z=9 containing a $+72^{\circ}$ disclination)
and (0,5,4)'s ($\sim$ 20\%).  These types are typical of the
tri-capped trigonal prism (TTP) and the mono-capped square archimedean
prism (a slightly distorted variant of TTP) respectively. The TTP is
found in the crystal structures of Fe$_3$B with Pearson symbols oP16
and tI32. The distorted (0,5,4) version is found in the structure
Fe$_{23}$B$_6$ of Pearson type cF116. These structures have been
identified as the leading competitors for B-Fe glass~\cite{Marek_Fe}.
Boron also took environments of the Kasper polyhedron (0,4,4) ($\sim$
3\%) corresponding to Z=8 and (0,2,8) ($\sim$ 10\%) corresponding to
Z=10. The association of B with $+72^{\circ}$ disclinations explains
how it increases the frustration of icosahedral order on the Fe atoms.
Clearly the improved glass-formability of FeB compared with elemental
Fe cannot be due to icosahedral order. Rather, it is presumed to be
caused by the deep eutectic at Fe$_{83}$B$_{17}$.

A Voronoi analysis of supercooled FeZrB at T=800K shows a clear
increase in the very negative \w6 polyhedrons, and also a decrease in
the number of Z14 (0,0,12,2) types on Fe atoms indicating a decrease
in frustration in the ternary as compared to the binary.  Environments
around B atoms were roughly similar in the binary and the ternary,
with a slight increase in the lower coordinated (Z=8) Kasper
polyhedrons at the cost of higher coordinated (Z=10) ones. Zirconium
took a variety of polyhedra, with an average coordination of 17.6 and
a minimal coordination of 15, owing to its large size compared to the
other constituents.

\subsection{\label{sec:Alloy_HA}Honeycutt and Andersen analysis} 

\begin{table}
\caption{\label{tab:HA_alloy}HA analysis for supercooled FeZrB at T=800K}
\begin{tabular}{|r|r|r|r|r|r|r|}
\hline
\multicolumn{7}{|c|}{Root Pair} \\
\hline
 & Fe-Fe & Fe-Zr & Fe-B & Zr-Zr & Zr-B & B-B\\
 & & & & & &\\
\hline
\bf 14 pairs &\bf 0.30 & \bf 0.19 &\bf 0.54 & \bf 0.11 &\bf 0.55  &\bf 0.75\\
\hline
142 pairs & 0.08 & 0.06 & 0.03 & 0.06 & 0.03  & 0.02\\
\hline
143 pairs & 0.17 & 0.11 & 0.35 & 0.04 & 0.33  & 0.33\\
\hline
144 pairs & 0.05 & 0.02 & 0.17 & 0.01 & 0.19  & 0.41\\
\hline
\hline
\bf 15 pairs  &\bf 0.56 &\bf 0.53 &\bf 0.39 &\bf 0.48 &\bf 0.41 &\bf 0.18 \\
\hline
154 pairs & 0.22 & 0.27 & 0.03 & 0.36 & 0.04 & 0.0 \\
\hline
155 pairs & 0.33 & 0.22 & 0.37 & 0.08 & 0.36 & 0.17\\
\hline
\hline
\bf 16 pairs &\bf 0.10 &\bf 0.24 &\bf 0.01 &\bf 0.36 &\bf 0.01 &\bf 0.01\\
\hline
166 pairs & 0.09 & 0.20 & 0.0 & 0.26 & 0.00 & 0.00\\
\hline
\end{tabular}
\end{table}

We made a HA analysis of the ternary glassy alloy, by looking at root
pairs of chemical species $\alpha$ and $\beta$, choosing $R_{cut}$ in
the manner of the \w6 analysis. The frequency is normalized to sum to
one for each species pair $\alpha\beta$. Table~\ref{tab:HA_alloy} lists
the fraction of different $1x$ pairs in supercooled FeZrB.

Among the $1x$ pairs with Fe as one of the root pairs, the 15's are
most abundant at all temperatures. The 15's are mainly comprised of
155's and the 154's, with the 155's being always higher than the
154's. The percentage of 15's is similar for FeZrB, FeB as well as
pure Fe. Note that the 15's are largest for the Fe-Fe pairs.  They
also show a steady enhancement with supercooling, unlike the 14's
which decrease with supercooling. Among the 14's, the
icosahedron-related 143's for all root pairs are always higher than
the close-packed 142's, and remain fairly constant with
supercooling. The 14's are maximal for FeB root pairs and minimal for
FeZr.

Supercooled pure Fe has a high percentage of ``16'''s (13\%) compared
to pure Cu (7\%). The high number of ``16'''s in Fe is related to the
occurance of Z14 (0,0,12,2) environments in which the 6-fold bond
carries a $-72^{\circ}$ disclination, rather than the BCC 14 atom
arrangement which would also show a high degree of four-fold and
six-fold bonds (section~\ref{sec:Vor}). Adding B to Fe, shows an
increase of ``16'''s for Fe-Fe pairs to 21\% in FeB (not
shown). Adding large Zr atoms to FeB decreases the occurance of 16's
on the Fe-Fe pairs, putting them on Fe-Zr pairs and easing the
frustration of Fe centers. This causes the geometry about Fe centers
to be more icosahedral, and hence the shift of the \w6 towards
negative values.

Among the root pairs not containing Fe, note that the B-B pair has the
maximum 14's ($+72^{\circ}$ disclinations), especially 144's, while
the Zr-Zr pair has the maximum 16's ($-72^{\circ}$ disclinations),
especially the 166's, emphasizing the role of size in controlling the
frustration in alloys.

\section{\label{sec:conclude} Conclusion}

This study quantifies icosahedral and polytetrahedral order in
supercooled liquid metals and alloys.  This is the first such analysis
of glass-forming Fe compounds using configurations from
first-principles simulations.  While the structural properties of Fe
and Cu strongly resemble each other at high temperature, and indeed
are close to a maximally random jammed structure~\cite{MRJ,Cupaper},
their behavior evolves substantially, and in different manners, as the
liquid is supercooled.  Proper modeling of atomic interactions is
essential to capture the differing behavior of each element, and use
of a first-principles simulation is the most reliable means of
achieving this.

For pure elements we find the degree of local icosahedral order in the
supercooled liquid depends on the low temperature crystal structure,
with BCC metals such as Fe and W accommodating icosahedra more readily
than the FCC element Cu.  Alloying with large or small atoms can
further influence the degree of icosahedral order, with small atoms
(e.g. B in Fe) aggravating the frustration by introducing positive
disclination line defects, while large atoms (e.g. Zr in Fe) naturally
stabilize negative disclination line defects, relieving frustration on
the medium-sized Fe atoms.  While the enhanced glass-forming ability
of FeB compared to Fe cannot be related to icosahedral order, we
suggest that the formation of icosahedral order and disclination line
network, together with the slow dynamics of chemical ordering in a
complex alloy and the destabilization of competing crystal
phases~\cite{Marek_Fe}, enhances the glass-forming ability of FeZrB
compared with FeB.

\begin{acknowledgements}
This work was supported in part by DARPA/ONR grant N00014-01-1-0961.

\end{acknowledgements}

\bibliography{ico}% Produces the bibliography via BibTeX.

\end{document}